\documentclass[journal=nalefd,manuscript=article]{achemso}

\usepackage[T1]{fontenc}       
\usepackage{graphicx,color}
\usepackage{amsmath}
\usepackage{amsfonts}
\usepackage{amssymb}
\usepackage[colorlinks=true,linkcolor=blue,citecolor=blue]{hyperref}

\author{Minkyung Jung}
\email{minkyung.jung@unibas.ch}
\altaffiliation{These authors contributed equally to this work.}
\affiliation[University of Basel]
{present address: Division of Nano-Energy, DGIST, 333 Techno Jungang-Daero, Hyeongpung, Gaegu, Korea 42988}

\author{Peter~Rickhaus}
\altaffiliation{These authors contributed equally to this work.}
\author{Simon~Zihlmann}
\author{Peter~Makk}
\author{Christian Sch{\"o}nenberger}
\email{christian.schoenenberger@unibas.ch}
\affiliation[University of Basel]
{Department of Physics, University of Basel, Klingel\-berg\-strasse 82, CH-4056 Basel, Switzerland}

\title{Microwave photodetection in an ultraclean suspended bilayer graphene pn junction}


\begin{document}

\begin{abstract}
We explore the potential of bilayer graphene as cryogenic microwave photo\-detector by studying the microwave absorption in fully suspended clean bilayer graphene p-n junctions in the frequency range of $1-5$\,GHz at a temperature of $8$\,K. We observe a distinct photo\-current signal if the device is gated into the p-n regime, while there is almost no signal for unipolar doping in either the n-n or p-p regimes. Most surprisingly, the photo\-current strongly peaks when one side of the junction is gated to the Dirac point (charge-neutrality point CNP), while the other remains in a highly doped state. This is different to previous results where optical radiation was used.
We propose a new mechanism based on the photo\-termal effect explaining the large signal. It requires contact doping and a distinctly different transport mechanism on both sides: one side of graphene is ballistic and the other diffusive. By engineering partially diffusive and partially ballistic devices, the photocurrent can drastically be enhanced.


\end{abstract}

\newpage

Graphene has shown a great number of exceptional electrical, mechanical and thermal properties.\cite{Sarma_Review2011,Neto_Review2009} Importantly, graphene is also a promising photonic material\cite{Bonaccorso_Nature_photo2010} whose gapless band structure allows electron-hole pairs to be generated over a broad energy spectrum, from ultraviolet to infrared.\cite{Koppens_Nature_Nano2014,Dawlaty_APL2008} In addition, photonic devices operate at high speed, due to the high mobility.\cite{Xia_Nature_nano2009,Mueller_Nature_photo2009} Recently, a number of novel applications have been considered in a variety of graphene photonic devices,\cite{Bonaccorso_Nature_photo2010} such as transparent electrodes in displays,\cite{Bae_Nature_Nano2010} tera\-hertz lasers\cite{Chakraborty_Science2016} and plasmonic systems\cite{Grigorenko_Nature_photo2012}.

Due to the flat wide-bandwidth photonic absorption of graphene, a significant effort has been made to demonstrate graphene photo\-detectors by measuring the photo\-current at graphene-metal contacts,\cite{Xia_Nature_nano2009, Lee_Nature_Nano2008,Park_NL2009,Mueller_Nat_photo2010,Nazin_Nat_Physics2010} monolayer-bilayer interfaces\cite{Xu_NL2009} and graphene p-n junctions.\cite{Song_NL2011,Gabor_Science_2011,Lemme_NL2011,Liu_ACS_Nano2012} There have been debates regarding the photocurrent generation mechanisms in graphene devices. In early studies with graphene-metal contacts,\cite{Xia_Nature_nano2009, Lee_Nature_Nano2008,Park_NL2009,Mueller_Nat_photo2010,Nazin_Nat_Physics2010} the interpretation of the photo\-current was based on the photo\-voltaic (PV) mechanism, in which a built-in electric field separates the photo\-generated charge carriers sufficiently fast so that electrons and holes do not recombine but rather add to a net photo\-current. More recent studies have demonstrated experimentally and theoretically that the photo\-thermo\-electric (PTE) effect is the dominant mechanism in graphene p-n junctions.\cite{Song_NL2011,Gabor_Science_2011,Lemme_NL2011,Xu_NL2009} The PTE effect arises from a light-induced temperature increase resulting in a thermo\-electric voltage. So far, graphene photo\-detectors have only been demonstrated for optical wavelengths, from near infrared to ultraviolet. Photo\-detection in the microwave range has not yet been studied although graphene has a very high potential as a fast detector.

In this work, we report on the observation of a microwave (MW) induced photo\-current in a fully suspended and ultra\-clean bilayer graphene p-n junction device. At zero source-drain bias a quite large current is measured in the bipolar region and an even more pronounced one when one side of the junction is gated to the charge-neutrality point (CNP), while the photo\-current is strongly suppressed in the unipolar region. This is in agreement with the notion that electron-hole pairs can only be generated when there is a region in the device where the Fermi energy is close to zero (at the CNP). Otherwise, photon absorption through direct band transitions is blocked by state occupancy (Pauli blockade). As we will demonstrate, the large photo\-current signal arises due to an asymmetry in transport properties, when the low doped side around the CNP becomes partially diffusive, while the other side remains highly doped and therefore ballistic. This will result in a temperature profile that maximize the photo\-termal signal.

\begin{figure}[t]
    \centering
        \includegraphics[width=0.7\textwidth]{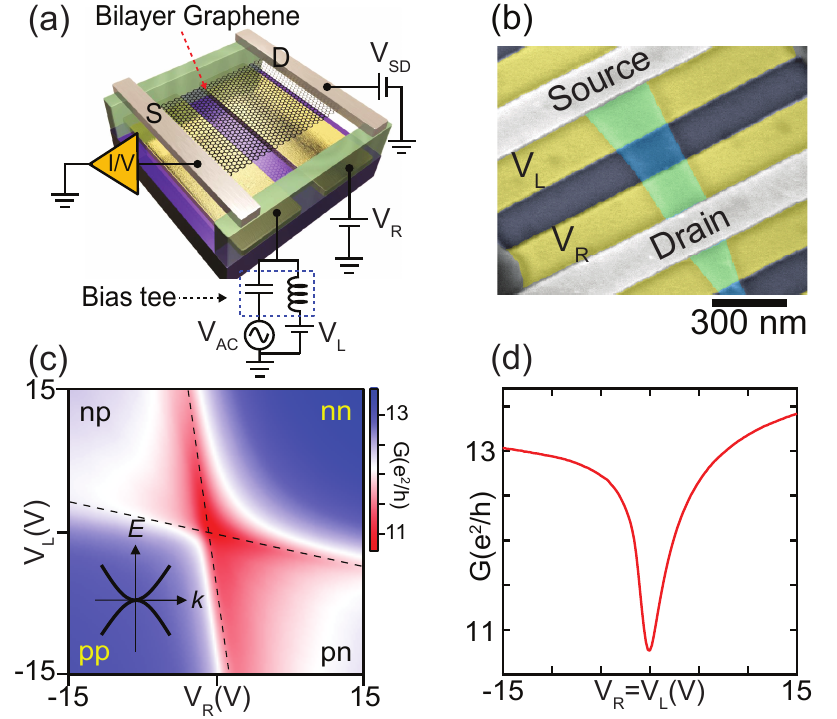}
    \caption{(Color online) (a) Schematics of the measurement setup and device geometry. The MW signal is applied to the left gate via a bias tee. (b) Scanning electron microscopy image of a suspended bilayer graphene device. (c) Conductance measured as a function of the left ($V_{L}$) and right bottom gates ($V_{R}$). Four regions are labeled according to carrier doping, $p$-type or $n$-type in the left and right regions controlled by the respective bottom gates. Inset:  quadratic energy-momentum dispersion relation of bilayer graphene. (d) Conductance trace measured for unipolar doping from $pp$ to $nn$ realized by setting $V_L=V_R$.}
\end{figure}


A device schematic and measurement setup is shown in Fig.~1(a).  The graphene device is suspended using a polymer based suspension method\cite{Tombros2011,Rickhaus_Nature2013,Maurand_Carbon2014} and the fabrication follows ref.~\cite{Maurand_Carbon2014}. In brief, an array of Ti/Au gate wires on a highly resistive oxidized Si substrate is defined first. The gates are $45$ nm thick, $600$ nm wide, and spaced at a $600$ nm pitch. After covering the bottom gate array with $600$ nm thick lift-off resist (LOR 5A, MicroChem Corp.), an exfoliated piece of graphene is transferred onto the LOR aligned to the gate array by using a mechanical transfer technique. Two Pd contacts spaced by $1.3$ $\mu$m are fabricated on the graphene (Fig.~1(a)). Finally, the LOR layer underneath the graphene flake is e-beam exposed and developed, suspending the graphene and source-drain contacts. Fig.~1 (b) shows a SEM image of a fully suspended graphene device (not the one measured). The actual device, whose measurement we report on, is wider than long. It has a width of $W=3.8$~$\mu$m and a length of $L=1.3$~$\mu$m.

Here, we use bilayer graphene which has a quadratic energy-momentum dispersion relation as shown in the inset of Fig.~1(c). Since the density of states of bilayer graphene is larger than that of single-layer in the vicinity of the CNP, electron-hole pairs can be generated more efficiently.  The device is then mounted and bonded to a circuit board on which both radio-frequency (RF) and DC lines are implemented. A MW signal $V_{AC}$ is coupled to the DC gate voltage $V_L$ of the left gate via a bias-tee (Fig.~1(a)). A source-drain DC bias voltage $V_{SD}$ is applied at the right graphene contact, while a current$-$voltage ($IV$) converter connected to ground is used to read out the DC current at the left contact. The device is measured in a cryostat in vacuum at a temperature of $\sim 8$ K. The as-fabricated device initially exhibits a weak gate dependence, indicating strong doping by resist residues. To remove these dopants and obtain ultra\-clean graphene, in-situ current annealing is performed.\cite{Rickhaus_Nature2013,Maurand_Carbon2014}

Figure 1(c) shows the electrical conductance $G$ versus $V_{L}$ and $V_{R}$ for the bilayer graphene p-n junction obtained by applying a source-drain voltage of $V_{SD}=400$ $\mu$V after current annealing. $G$ exhibits four characteristic regions p-p, n-n, p-n and n-p (the first symbol refers to the left and the second to the right region) according to carrier doping in the left and right regions depending on both gate voltages. The border lines for the four different regions are close to perpendicular to each other, indicating that capacitive cross coupling between the two gates is weak. Figure 1(d) shows $G$ for unipolar doping from the $pp$ to the $nn$ regime where $V_L = V_R$ holds. The pronounced dip in $G$ signals the CNP (Dirac point) with the minimum appearing at a gate voltage close to zero reflects the absence of uncontrolled doping. It is seen that conductance already starts to saturate for large gate voltages due to contact resistances, evaluated in the supplementary. The asymmetry in saturation indicates n-type doping of the graphene below the metal contacts. This is consistent with our previous work.\cite{Rickhaus_Nature2013}


\begin{figure}[t]
    \centering
    \includegraphics[width=0.7\textwidth]{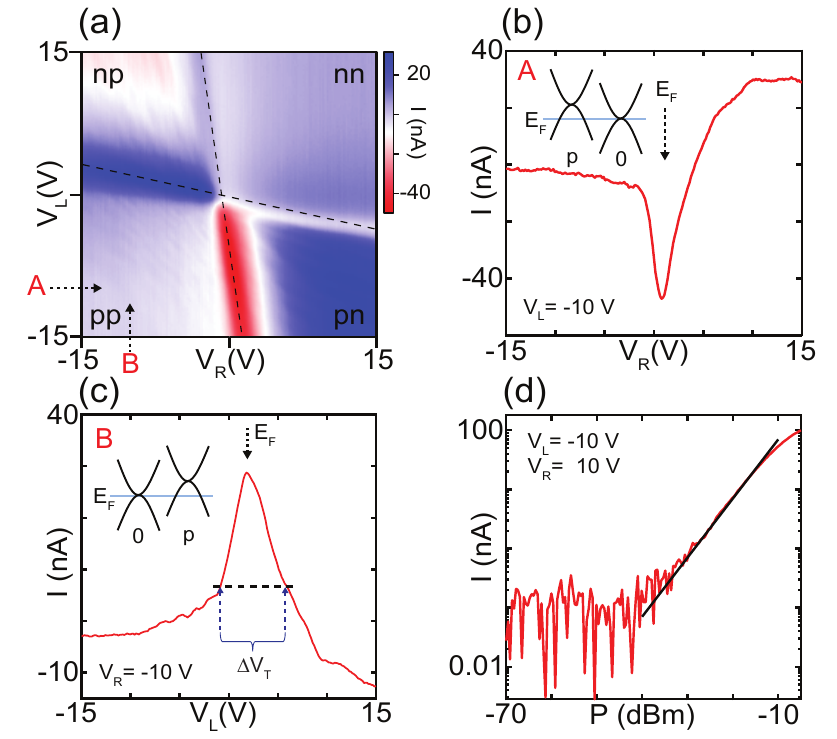}
    \caption{(Color online) (a) Measured DC photo\-current $I$ as a function of $V_{L}$ and $V_{R}$ at $V_{SD}$ = 0 V while applying an MW signal at $f$ = 3.81 GHz with a power of $P_{RF} = -14$ dBm. Dashed lines locate the CNPs in the left and right regions taken from the conductance measurement in Fig.~1(c). (b,c) Photo\-current line traces taken for $V_L=-10$~V (trace A) and $V_R=-10$~V (trace B). Dashed arrows in (b) and (c) indicate the CNP. (d) shows the dependence of the photo\-current on the MW power $P$ applied at room temperature measured in the bipolar p-n regime for $V_L=-10$ and $V_R=10$~V in log-log representation.\cite{comment1} The black line corresponds to $I\propto P$.}
\end{figure}

We next look into the photo\-current experiment. This is done by setting $V_{SD}=0$~V and applying an AC MW signal to the left gate via a bias-tee as shown in Fig.~1(a). Figure 2(a) shows the map of the measured DC photo\-current $I$ as a function of $V_{L}$ and $V_{R}$ for a MW signal with frequency $f = 3.81$~GHz and power $P = -14$~dBm.\cite{comment1}
%
We observe qualitatively similar results at different frequencies $f\gtrsim 1$~GHz but with different current values (Fig.~S1(a), supplementary data). Importantly, without applying a MW signal, the photo\-current pattern vanishes. There is only a background noise signal left amounting to $\sim$ 50 pA, which is independent of the gate voltages. This is shown in Fig.~S1(b) (supplementary data). The dashed lines in Fig.~2(a) mark the CNPs in the left and right regions and are taken from the $G$ measurement in Fig.~1(c).

In discussing the photo\-current pattern $I(V_L,V_R)$ we first observe a very low signal in the unipolar regime. In contrast, a pronounced photo\-current appears in the two bipolar regions p-n and n-p with opposite signs. Most remarkably, the most intense photo\-current is observed along the CNPs indicated by the black dashed line. The signal is particularly large for the case p-0 and 0-p, again with opposite sign. Figure 2 (b) and (c) show line traces $I(V_R)$ taken at $V_L=-10$~V and $I(V_L)$ at $V_R=-10$~V, corresponding to cut A and B indicated by arrows in Fig.~2 (a). The intense photo\-current is particularly well visible in these two cuts appearing as a pronounced dip and peak in (b) and (c) at the respective CNP. The dependence of $I$ on the MW power $P$ applied at room temperature is plotted in Fig.~2(d).\cite{comment1} It has been measured in the bipolar p-n region with $V_{L} = -10$~V and $V_{R} = 10$~V. The photo\-current starts to exceed the background at around $P=-40$~dBm and then grows approximately linear with $P$ (black line).

 Since a p-n device, for which the p and n regions have the same size and are contacted by the same material, is mirror symmetric, we expect $I(V_L,V_R) = -I(V_R,V_L)$ to hold, i.e. exchanging left with right reverts the sign of $I$. Qualitatively, this symmetry is quite nicely present in the measured data and seen in Fig.~2(a-c). However, close inspection shows that the symmetry is not exact. We think that this is due to an asymmetry in either the gate pattern or doping in the contacts breaking the mirror symmetry. If the photo\-current is due to the PV effect, one would expect the largest signal in the middle of the bipolar regions where $p=-n$. In addition, a single sign change is expected at the line $V_L=V_R$ in the unipolar regime where $I=0$ (again due to symmetry reasons). Multiple sign changes (or multiple minima and maxima) in the photo\-current have been observed in the literature and taken as evidence that another effect is causing the signal. Since previous data measured for optical light could very well be described by the PTE model,\cite{Song_NL2011,Gabor_Science_2011,Lemme_NL2011,Xu_NL2009} we follow along the same route. Although our data looks different in some important aspects to previous experiments, in particular the very intense photo\-current at the CNPs is a new observation, there are multiple maxima and minima in $I$ present, suggesting a PTE origin.

Next, we will compare the measured photo\-current with the PTE effect using a simple one-dimensional continuum model without contact doping.\cite{Song_NL2011}  We assume that we can neglect the PV effect and that there is a mechanism by which part of the MW signal is absorbed and dissipated in the electron gas of the graphene layer leading to an effective local electron temperature $T$ larger than the measurement bath temperature $T_0$ of the cryostat. Gradients in $T$ will produce a thermoelectric voltage via the thermo\-power $S$, also known as the Seebeck coefficient. The open-circuit photo\-voltage $V$ can then be written as
\begin{equation}
   V=\int S dT = \int_0^L S(x)\frac{\partial T}{\partial x}dx \quad \text{.}
\end{equation}
The graphene contacts are located at position $x=0$ and $x=L$ and the integral is taken from contact to contact with the boundary condition $T(0)=T(L)=T_0$. In assuming local equilibrium, $S(x)$ will be given by the carrier density $n(x)$ at position $x$, i.e. $S(x) = S(n(x))$. If we now assume that the intrinsic part of the graphene device can be separated into a left and right region having constant doping and a narrow p-n interface, $S$ is constant in the left and right region with values $S_L$ and $S_R$. We therefore obtain for $I$ the simple result
\begin{equation}
   I=G \left(S_L-S_R\right) \Delta T \quad \text{.}
\end{equation}
Here, $G$ is the device conductance and $\Delta T=T(L/2)-T_0$, assuming $x=L/2$ to be the location of the p-n interface.
Assuming further diffusive transport, the Seebeck coefficient of the homogeneously doped left and right graphene regions can be expressed as\cite{Hwang_PRB2009}
\begin{equation}
  S_i = -\frac{\pi^2k_B^2T}{3e}\frac{1}{G_i}\frac{\partial G_i}{\partial \mu_i} \quad \text{,}
\end{equation}
where $i$ refers to the left or right region, i.e. $i=L,R$, $k_{B}$ is the Boltzmann constant and $\mu_i$ the chemical potential in region $i$. The dependence of $S$ on the gate voltages $V_i$ can be obtained by writing $\partial G_i/\partial \mu_i$ as $(\partial G_i/ \partial V_i)(\partial V_i / \partial \mu_i)$. For bilayer graphene, the second term is a constant given by $2m/(\hbar^2 \pi C'_g)$, where $m$ is the bilayer mass ($m=0.03$\,$m_e$) and $C'_g$ the gate capacitance per unit area. In order to continue, we need a model for the total $G$ of a p-n device having two different regions. Following Song~{\it et al.}\cite{Song_NL2011} we take a classical resistor model and write $G^{-1}=G_L^{-1}+G_R^{-1}$, where $G_L$ and $G_R$ are the graphene conductances of the left and right part separately. To obtain $S_i$ we describe $G_i$ in a phenomenological manner as
\begin{equation}
  G_i(\mu_i) = G_{min}\sqrt{1 + \left(\mu_i/\Delta\right)^2} \quad \text{,}
\end{equation}
where $G_{min}$ is the minimum of the conductance at the CNP and $\Delta$ models the width of the conductance dip around the CNP. The functional dependence has been chosen such that $G_{L,R}(\mu_{L,R})\propto \mu_{L,R}$ for a large potential as required for bilayer graphene. Taking all together one can calculate the expected photo\-current $I(V_L,V_R)$ as a function of the two gate voltages $V_{L,R}$ for a constant (yet unknown) temperature difference. The result is shown in Fig.~3(a) and a corresponding plausible temperature profile is shown in Fig.~3(c). Before discussing this result further, let us mention the limitations of the model right away. The model assumes classical diffusive transport with a well defined local potential and local thermal equlibrium. We have further neglected the effect of the graphene band-structure on the resistance of a p-n junction (Klein tunneling). The resistance in the bipolar region is definitely larger than in the unipolar regime even for the same (absolute) doping. For this classical model $G(V,-V)= G(V,V)$, while in reality $G(V,-V)< G(V,V)$. Since our sample displays a pronounced conductance minimum with almost zero residual extrinsic doping (CNP close to zero gate voltage), we expect ballistic transport features to appear.\cite{Rickhaus_Nature2013} Fabry-Perot resonances are indeed visible in the thermoelectric signal (Fig.~S2, supplementary). Hence, this simple model we discuss here, can serve as a guide, but it is not expected to fit the data in full.

\begin{figure}[t]
    \centering
    \includegraphics[width=0.7\textwidth]{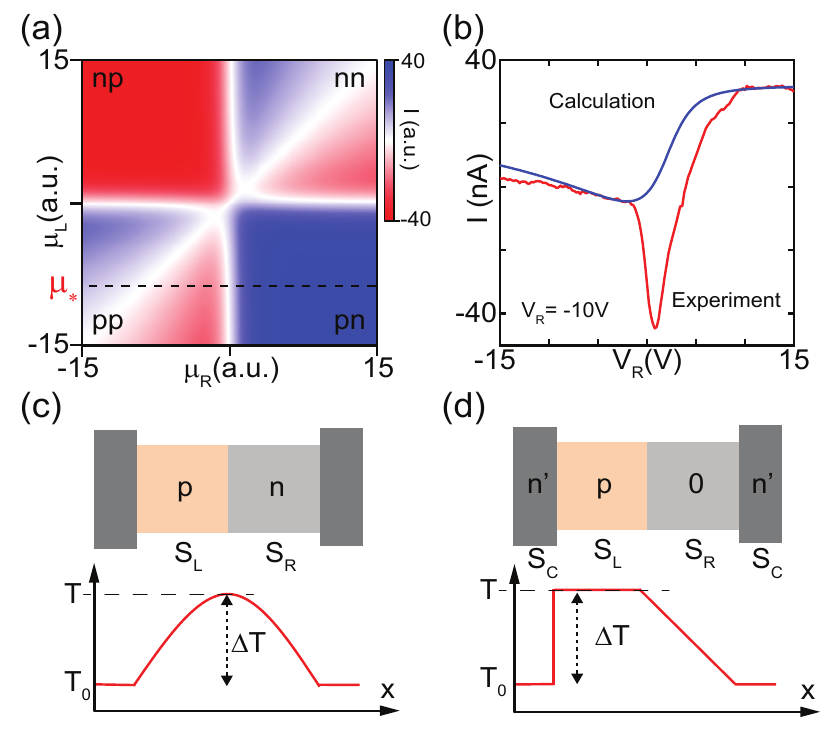}
    \caption{(Color online) (a) Calculated photocurrent map assuming a photo\-thermo\-electric (PTE) origin as a function of the chemical potentials $\mu_{L}$ and $\mu_{R}$ for a p-n junction. (b) Comparison between the calculation and experimental data taken at $\mu_{*}$ and $V_L=-10$~V, respectively. (c) Illustration of the temperature profile for the p-n region (where $p=-n$) without contact doping and assuming ideal thermalization at the graphene contact interface. $S_{L}$ and $S_{R}$ denote the Seebeck coefficients in the left and right gate regions and $\Delta T$ the maximal temperature increase relative to the surrounding. (d) Illustration of an effective temperature profile for the $p0$ case (p-doped on the left side and at the CNP on the right), including the contact doping $n^*$. $S_{c}$ denotes the Seebeck coefficient at contacts.}
\end{figure}

As shown in Fig.~3(a), the calculated photo\-current $I$ displays multiple sign changes which arise from the difference $\Delta = S_{L}(V_L)-S_{R}(V_R)$ of the left and right Seebeck terms. The expected symmetry $I(V_L,V_R)=-I(V_R,V_L)$ is nicely respected. While the current is positive in the p-n regime, it is negative in the n-p one, but there is also a photo\-current in the unipolar quadrants, providing that the doping is not the same in the two sides. There are major differences immediately apparent when comparing the experimental data in Fig.~2(a) with the model in Fig.~3(a). While there are multiple sign changes in the experimental data as well, the most pronounced signal is not concentrated in the middle of the bipolar region, but rather close to the Dirac point. As we have already pointed out, a very strong signal appears for the two doping states 0-p and p-0, when one region is close to the CNP and the other p-doped. However, if we neglect this very strongly peaked signal for the moment, the experiment compares quite well with the model. This can best be seen, when comparing cross-sections. A line profile taken at $\mu_{L} = \mu_{*}$ in Fig.~3(a) is presented with the experimental data in Fig.~3(b) taken at $V_L=-10$~V. The model describes very well the photo\-current in the p-p and p-n regions including the transition when neglecting the strongly peaked (negative) signal at the CNP. Hence, the background can surprisingly well be described by a simple classical resistor model. The strong dip at the CNP, however, must have a different origin. First, we repeat that the signal is strongest for doping configurations 0-p and p-0. There is also an enhanced signal for the configurations 0-n and n-0, but it is smaller in magnitude by at least a factor of $2$. The device therefore breaks charge conjugation symmetry within the device. This can be explained by contact doping which we have neglected in the description of $I$ until now. We have mentioned in the beginning that the graphene source and drain contacts induce an n-type doping in the contact areas. This is evidenced by the asymmetric shape of $G$ for large gate voltages in a unipolar gating configuration, Fig.~1(d). Let us denote the contact doping by $n_C$. Including the contacts a certain doping configuration of the device would then read n$_C$-p-n-n$_C$, for example. Charge conjugation is now no longer an internal symmetry of the device. This points to the important role of the contacts in understanding the large photo\-current signal, which we observe here.

If we stay within a model with local thermal equilibrium and a description following the PTE effect including now the graphene contacts described by graphene regions with a fixed contact doping $n_C$, the temperature profile $T(x)$ shown in Fig.~3(d) can explain the large signal for the internal doping state p-0. As compared to the the case without contact doping shown in Fig.~3(c), the temperature profile is very asymmetric. The explanation is as follows: First, we assume that both contacts are ideal heat sinks so that the temperature is fixed to $T_0$ there. Second, the fact that a much larger photo\-current appears when one side of the graphene p-n junction is gated into the CNP suggests that MW absorption is very effective in the zero-doped regions. Consequently, in the case shown in (c) most of the hot carriers are generated in the middle at the interface of the p-n junction. The carriers then diffuse symmetrically to the left and right, disposing the energy to both contacts. This yields the temperature profile shown in (c) used to model the photo\-current before. If we take contact doping into account and consider the doping state n$_C$-p-0-n$_C$, MW absorption will mainly take place in the right 0-region. The motion of electrons at the CNP can be described as quasi-diffusive due to the random puddle landscape. Although the carrier density is minimal here, heat can diffuse quite well to the right contact yielding a temperature gradient as indicated in the figure. On the left side, however, a quite strong p-n junction is expected in the vicinity of the contact. Since remaining inhomogeneities are well screened at high carrier doping, electron propagation is expected to be ballistic in the left $p$-region. This is confirmed by the observation of Fabry-Perot resonances (Fig.~S2, supplementary).\cite{Rickhaus_Nature2013} Due to the p-n junction at the contact, many of the hot electrons that propagate oblique towards the junction will be reflected. This results in a relatively large thermal resistance causing the temperature drop in the left contact as shown in (d). The expected photo\-current would then be given by $I=GS_C \Delta T$, if we assume that the temperature drops fully within the contact doped region and because $S_R=0$ as the right regions is at the CNP.

 Since $S_C < 0$, this yields a negative photo\-current which is largely independent of $V_L$ in agreement with the experiment. It is clear that this picture is simplified and it would be interesting to describe the hot carrier distribution in a refined (quasi-) ballistic model.
Using this equation, $I=G S_C\Delta T$, we estimate a temperature rise $\Delta T$ of $\sim 1-2$\,K taking a typical photo\-current of $30$\,nA at a gate voltage of $V_L=5$\,V and $V_R=0$\,V corresponding to a chemical potential of $\sim 2$\,meV and using equ.~3 to estimate $S_C$.

\begin{figure}[ht]
    \centering
    \includegraphics[width=0.6\textwidth]{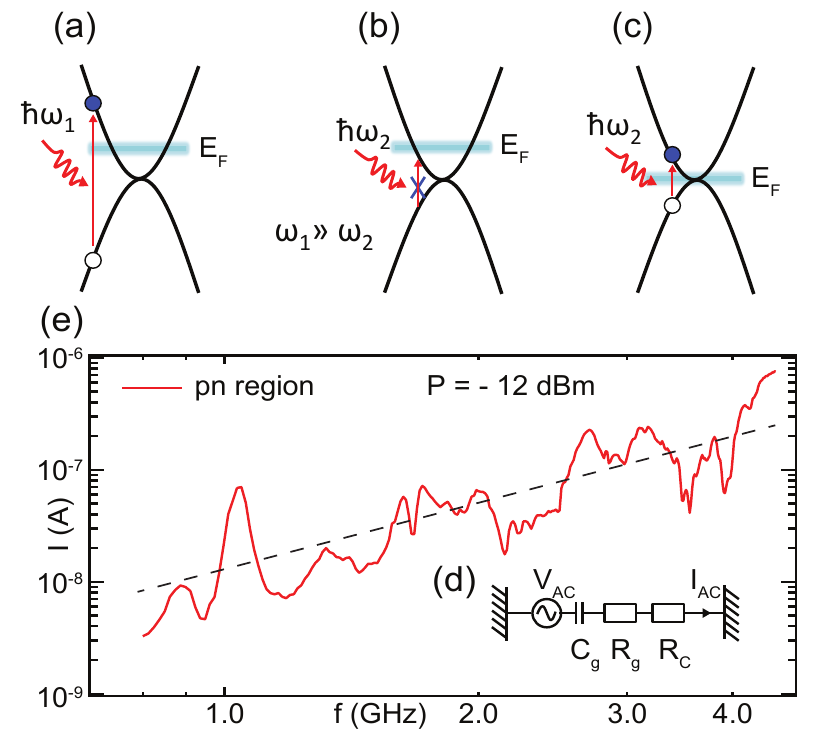}
    \caption{(Color online) (a-b) Illustrations of direct optical absorption processes in doped bilayer graphene for different photon energies $\hbar \omega$  relative to the chemical potential $E_F$. For $\hbar\omega_{1} > 2 |E_{F}|$, shown in (a), the process is allowed, whereas for $\hbar\omega_{1} < 2 |E_{F}|$, shown in (b), the process is not possible due to Pauli blocking. In (c) we indicate the situation at the CNP for zero doping. Direct optical transitions are allowed also for small photon energies $\hbar\omega_2$. (d) inset to (e): effective circuit diagram relevant for the MW gate modulation. $C_g$ denotes the gate capacitance and the graphene resistance is split into two parts, the contact resistance $R_C$ and an intrinsic graphene resistance $R_g$. (e) Photo-current in the p-n regime at gate voltages $(V_L,V_R) = (-14,14)$~V measured as a function of MW frequency $f$ at a power of $P = -12$~dBm. The frequency-dependent attenuation in the RF cable has been corrected. The effective power at the sample is in this graph $P\approx -17$\,dBm. The dashed line corresponds to slope $2$.}
    \label{fig:4}
\end{figure}

We see in the data that the largest photo-current appears when one side of the p-n junction is gated to the CNP. This suggests that MW absorption preferentially takes place in the zero-doped region. We can think of two reasons leading to this dependence: one based on the picture of direct optical transitions and the other on current-driven losses in graphene. In the following, we briefly discuss both.

Direct optical transitions generate electron-hole pairs and are only possible between filled and empty states as shown in Fig.~4(a-c). If one is using visible optical light with frequency $\hbar\omega_1 \approx 2$~eV, as done in other works,\cite{Koppens_Nature_Nano2014} $\hbar\omega_1 \gg 2|E_F|$ ($E_F$ is the gate-controlled chemical potential) even for large p or n-doped graphene. Electron-hole pair generation is then always possible, also if graphene is doped, Fig.~4(a). This is different, if a small frequency $\omega_2$ is used as shown in Fig.~4(b). If we consider MWs, a $10$~GHz signal corresponds to a photon energy of $40$~$\mu$eV which equals $\sim 0.5$~K. To compare with a typical doping state in our bilayer graphene device we estimate $E_F$ for a gate voltage of $V_g=10$~V. Using $E_F=\hbar^2\pi n/2 m$, where $n$ is the carrier density, $n=V_gC'_g/e$ with $C'_g\approx 15$~aF/$\mu$m$^2$ the gate capacitance per unit area. This yields $n\approx 10^{11}$~cm$^{-2}$ and $E_F\approx 4$~meV at $V_g=10$~V. It is obvious now that for any typical doping concentration in our device $\hbar\omega \ll E_F$. Hence, photo-absorption by a direct transition from the valence to the conductance band is forbidden by the Pauli principle, as either the relevant states are fully occupied or fully empty, Fig.~4(b). This is different, if there is a region in the device where the doping state corresponds to the CNP, hence, where $E_F=0$. This is the case, for example, in the bipolar regime, but not in the unipolar one. We have to take into account, however, that the experiment has not been conducted at zero temperature but at $T_0=8$~K. Since $k_BT_0 \gg \hbar\omega$ for our MW frequencies the picture in Fig.~4(c) for $E_F=0$ needs to be slightly refined. Due to the finite temperature the states in the valence and conduction band are occupied to $\approx 50$~\% within an energy bandwidth of $\approx k_B T_0$. Photon absoprtion is now allowed also for doped graphene as long as $E_F$ remains within a window of order $k_B T_0$ around the CNP. Once $|E_F| \gg k_B T_0 \gg \hbar \omega$ photon absoprtion of MWs must be blocked. Note, the residual doping in this graphene device is $5.5*10^9$\,cm$^{-2}$, which in energy corresponds to $~220$\,$\mu e$V (Fig.~S3, supplementary), which is much smaller than $T_0$.

The second picture uses an electronic-circuit description shown in Fig.~4(d). Modulating the gate voltage with the MW signal results in an AC current. The graphene sheet is charged and discharged in a periodic manner. The effective circuit is an $RC$ one where $C$ is given by the gate capacitance $C_g$ and $R$ by (parts of) the graphene resistance. Since the impedance of $C_g$ is much larger than $R$ for all frequencies of interest here, the ac current is directly given by $i\omega C_g V_{AC}$. Hence, the dissipated power is proportional to $R$ and to the frequency squared. Since $R$ is maximal at the CNP, dissipation is expected to be largest there. As seen in Fig.~1(d), $R$ only increases by $\approx 20$\,\% at the CNP, which would give rise to only a modest higher absorption at the CNP as compared to the highly doped case.
Since, this does not agree with the observation, we split $R$ into two parts: $R=R_C + R_g$,  where $R_C$ is the contact resistance, and $R_g$ is the intrinsic graphene resistance which is strongly peaked at the CNP. While the contact resistance is different for n and p doping, after substraction the intrinsic conductance displays a large change, reaching up to $10^3 \cdot G_0$ on the p side (Fig.~S3, supplementary). As a consequence, $R_g$ is strongly peaked around the CNP in agreement with the observed photo\-current signal.


We can distinguish between the two mechanism by studying the frequency dependence of the photo\-current $I$, see Fig.~4(e). In the case of MW photon absorption by direct transitions from the valence to the conductance band, the number of states that can participate in this process for a fixed frequency is proportional to the wave\-vektor $k$, where $k\propto \sqrt{\omega}$. The latter follows from the condition for absorption at the CNP given by $E=\hbar k^2/2m = \hbar \omega/2$. Hence, we expect that $I\propto \sqrt{f}$. If instead we describe the power absorption in a classical way using a resistor model, the MW drive would induce an AC current, which - as we already mentioned - is proportional to $f$. In our description the photo\-current $I$ is proportional to the temperature increase $\Delta T$ due to the absorbed power $P_{absorbed}$ intrinsic to graphene. Since, $\Delta T \propto P_{absorbed}$ and $P_{absorbed} \propto R_g I_{AC}^2$, we arrive at $I \propto f^2$. The measured $I(f)$ is shown in Fig.~4(e) on a log-log scale. $I$ increases with $f$ and displays a random pattern with pronounced peaks and dips. These features are not due to the graphene device but are caused by standing wave patterns arising in the sample box. It is clear that the $I(f)$ dependence is much closer to $\propto f^2$ than $\propto \sqrt{f}$ when comparing with the dashed line corresponding to slope $2$.
We therefore think that the dissipation mechanism can be captured by a resistor model, where the main dissipation leading to the temperature rise is not governed by the measured DC resistance, but by the intrinsic graphene resistance $R_g$. This resistance strongly peaks at the CNP, where the hot-carrier relaxation is particularly effective due to allowed electron-hole pair generation. This latter point suggests that Pauli blockade does also play a role in the second mechanism as it affects the relative magnitude in the microscopic relaxation processes which involve energy exchange between electrons alone or between electrons and holes as well-

A test for the Pauli-blockade of electron-hole pair generation is found in the width of the strong photo\-current peak seen in Fig.~2(a). We look at the cross section indicated by the arrow B and displayed as a graph in Fig.~2(c). As a function of gate voltage $V_L$ the doping configuration evolves from p-p to 0-p to n-p. If $|E_F|$ in the left regions gets larger than $\sim 4k_BT_0$, Pauli blockade should set in suppressing the signal. Hence, the width of the peak should be determined by the temperature $T_0$. Taking the appropriate gate-conversion factor $4k_BT_0$ is indicated in the graph by $\Delta V_T$. Hence, the width indeed conforms with the temperature. Although we did not do temperature dependent measurements, the peak is no longer discernable at large temperatures due to its increased width. This has been seen in a control measurement done at $T\approx 77$~K.

In conclusion, we report MW photo\-detection in a fully suspended and clean graphene p-n junction. By radiating a MW signal to bilayer graphene, a strong photo\-current signal is observed at zero bias in the bipolar regime, while the current is suppressed in the unipolar one. Most remarkably, the current is strongly enhanced when one of the two sides is at the CNP. The frequency dependence of the absorption signal favours a mechanism in which the graphene electron system is heated by a dissipation mechanism described by an intrinsic graphene resistance that strongly peaks at the CNP where hot electron carriers can relax by electron hole pair generation, while this mechanism is forbidden by Pauli blockade at larger doping. Our interpretation of the observed polarities of photo\-voltage or photo\-current is based on the photo\-thermo\-electric (PTE) effect. The largest signal, obtained when one side is at the CNP and the other side still highly doped, originates from two effects: the strong energy absorption in the low-doped diffusive region and a large temperature drop that builds up between the contact and the highly doped and ballistic region. The large temperature drop is caused by a large thermal resistance at this contact due to filtering of electron trajectories in this ballistic part. In the future, it would be interesting to model the MW-induced photocurrent in a self-consistent ballistic model taking both PV and PTE effects into account. It would also be interesting to measure the thermal resistance of a pn junction due to Klein tunneling.


\begin{acknowledgement}
We thank A. Baumgartner and M. Weiss for helpful discussions. This work was supported by the ERC project QUEST, the EC flagship Graphene, the Swiss National Science Foundation (SNF), including the project NCCR QSIT and the Swiss Nanoscience Institute.
\end{acknowledgement}


\begin{thebibliography}{99}
\bibitem{comment1}
\bibinfo{title}{Note, that $P$ is the power measured at the rf source output at room temperature. The power at the sample is lower due to damping in the cables. The reduction is frequency dependent, e.g. $-8$\,dB at $1$\,GHz and $-13$\,dB at $3$\,GHz.}
\end{thebibliography}
\end{document}


%


\newpage

\begin{figure}[htb]
  \begin{center}
    \includegraphics[width=0.9\columnwidth]{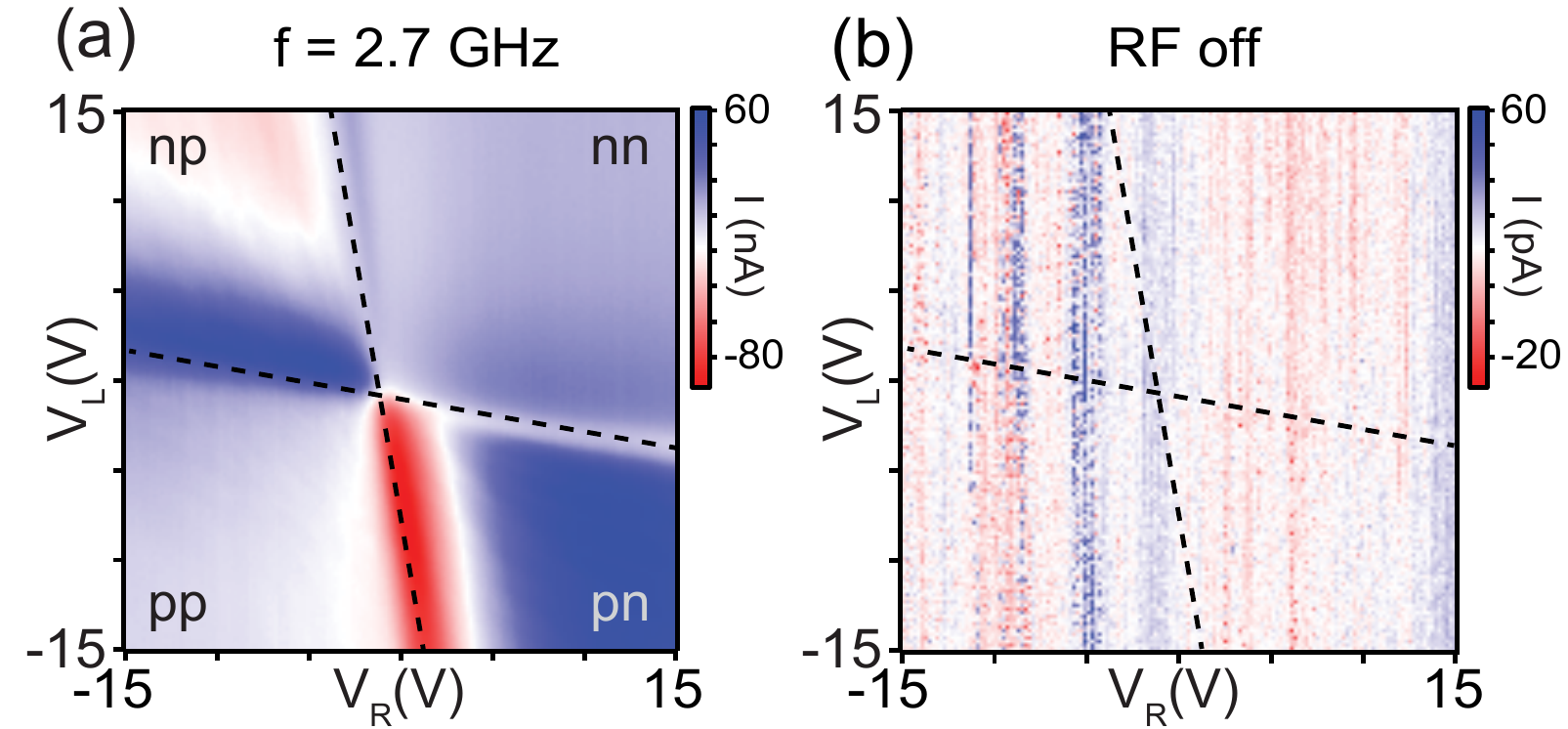}
  \end{center}
  \label{fig1}
\end{figure}
\noindent {\bf Figure S1}:
(a) Photocurrent measured as a function of $V_{L}$ and $V_{R}$ at $V_{SD}$ = 0 V, $f_{RF}$ = 2.7 GHz and $P_{RF}$ = -14 dBm. (b) Photocurrent measured as a function of $V_{L}$ and $V_{R}$ at $V_{SD}$ = 0 V and RF off. No current pattern is observed.


Figure~S1 is shows that the photocurrent is absent without any applied microwave signal. (a) shows the case RF on and (b) the case RF off.

\newpage

\begin{figure}[htb]
	\begin{center}
				\includegraphics[width=0.8\textwidth]{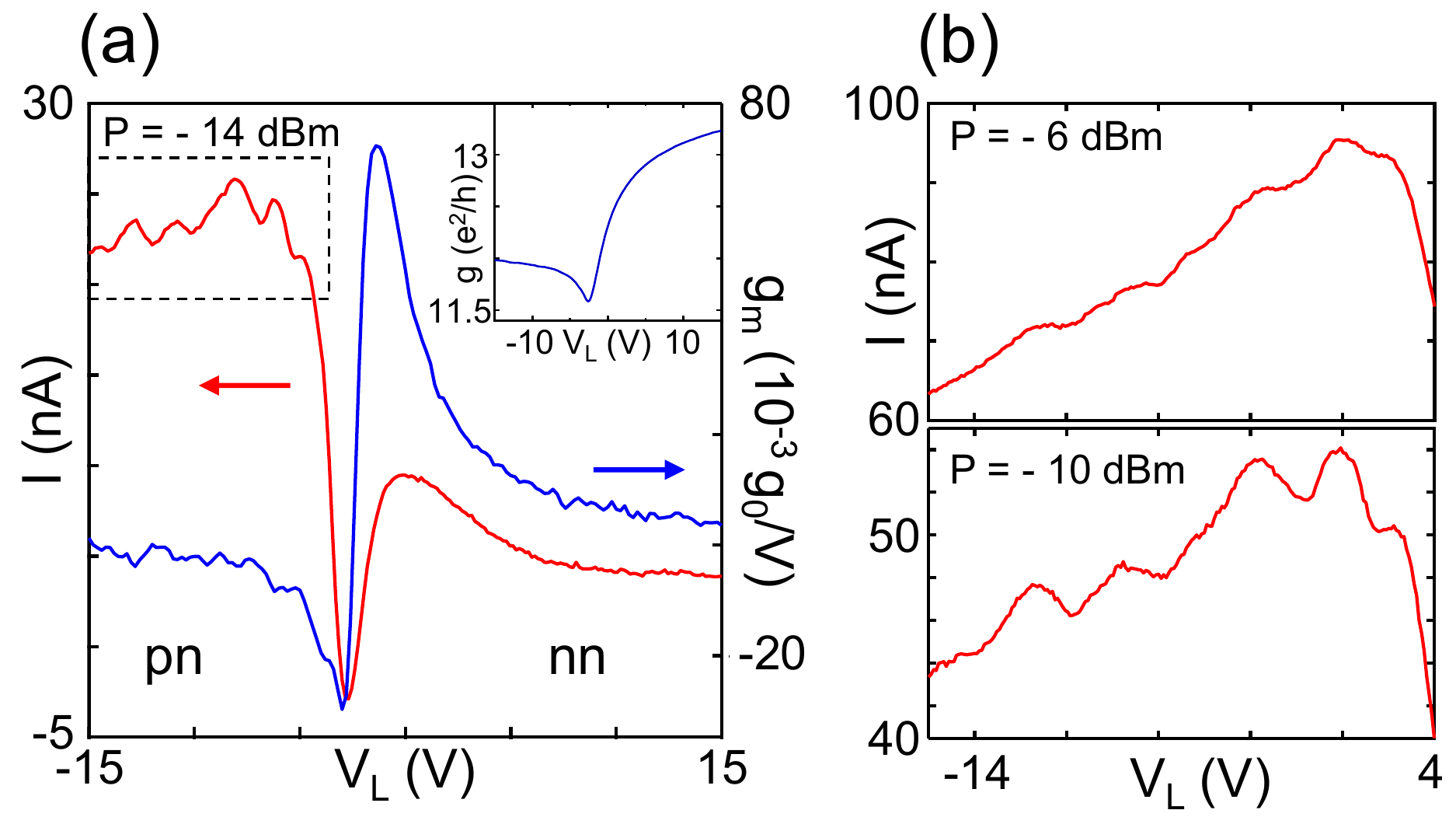}
	\end{center}
	\label{fig1}
\end{figure}
\noindent {\bf Figure S2}:
(a) Photocurrent (red curve) at $f=3.8$\,GHz and numerical trans\-conductance (blue curve) plotted as a function of $V_{L}$ at $V_{R}$ = 8.1 V. Inset shows the conductance measured at the same voltage range. (b) RF power dependence of the Fabry-P\'{e}rot (FP) oscillations. \\


We observe a clear Fabry-P\'{e}rot (FP) interference pattern in the photo\-current response in the suspended clean bilayer graphene device. The photocurrent and numerical trans\-conductance measured as a function of $V_{L}$ at $V_{R} = 8.1$\,V are plotted in Fig.~S2(a). The FP interference is not visible in the conductance as shown in the inset of Fig. S2(a) and can be barely resolved in the trans\-conductance curve. However, the interference pattern is clearly seen in the photo\-current plotted as a red curve in Fig.~S2(a,b).

The textbook FP resonance condition can be written as $j\lambda_j/2=L_C$, where $L_C$ is the cavity length, $\lambda_j$ the wavelength and $j$ the mode index with $j=1$ corresponding to the first eigen\-mode. The wavelength can be expressed by the carrier densities $n$, $\lambda = \sqrt{\pi/n}$. This then yields $n_j=\pi (j/L)^2$. If we assume a large mode index and calculate $\Delta n:=n_{j+1}-n_j$, we obtain $\Delta n \approx 2\sqrt{\pi n}/L_C$, where $n$ is the mean carrier density of the cavity. Using this equation for the higher mode spacing seen in Fig.~S2(b) yields a cavity length of $\sim 550$\,nm close to the geometrical length of $\sim 650$\,nm. We point out, however, that the simple equation predicts an oscillation spacing in carrier density, and therefore also in gate voltage, proportional to $j^2$. Hence, the spacing should rapidly grown with mode index $j$, which is not seen in the data. We think the reason that this is not seen is related to the cavity length which also changes with gate voltage. The FP spacing in gate voltage $V_L$ is almost constant in the measurement showing that the cavity length growths roughly like $L_C\propto \sqrt{n}$. Obviously, for large carrier density, when screening is ideal, the cavity length should asymptotically approach the geometric length of  $\sim 650$\,nm.

Figure~S2(b) shows the MW power dependence of the FP interference measured in the same area marked by the dashed box (a). The interference washes out when increasing the RF power to $P=-6$\,dBm. We estimate the temperature increase to $1-2$, $2.5-5$ and $6-12$\,K for the three cases $P=-14$, $-10$ and $-6$\,dBm, respectively.

\newpage

\begin{figure}[htb]
	\begin{center}
				\includegraphics[width=1.0\textwidth]{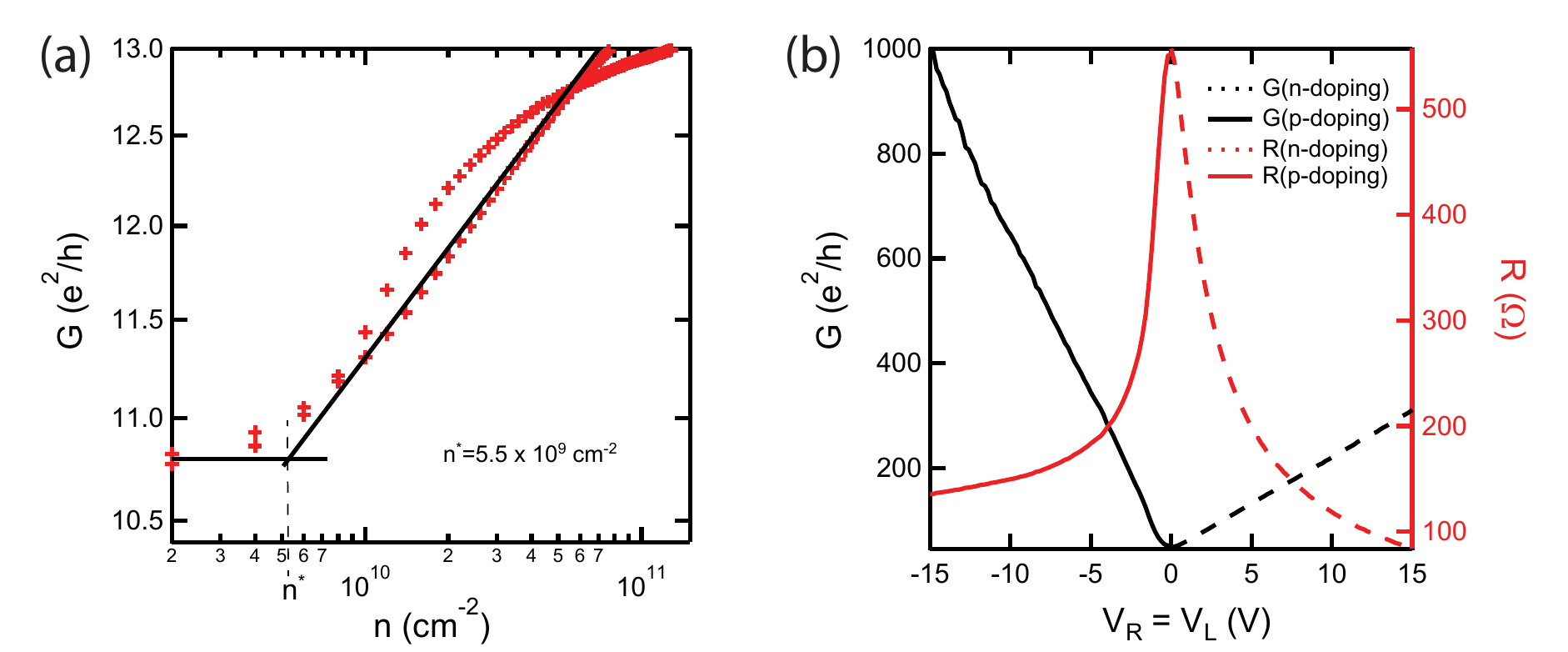}
	\end{center}
	\label{fig1}
\end{figure}
\noindent {\bf Figure S3}:
(a) Conductance $G$ for unipolar doping $V_L = V_R$ plotted as a function of the logarithm of the carrier density $n$. A crossover density of $n^*=5.5$\,cm$^{-2}$ is found, corresponding to an energy scale of $0.2$\,meV. (b) $G(V_L=V_R)$ after subtraction of a contact resistance $R_C$ and corresponding intrinsic graphene resistance $R$. The contact resistances are different for the two side. For the n-side $R_C=1.8$\,k$\Omega$ and for the p-side $R_C=2.1$\,k$\Omega$.\\


Although the dip in conductance $G$ at the charge neutrality point appears modest in Fig.~1(d) of the main text, this does not indicate a low quality as we demonstrate with Fig.~S3. $G(n)$ plotted as a function of unipolar carrier density $n$ yields a quite low crossover density of $n^*=5.5$\,cm$^{-2}$, supporting that a high quality and clean graphene device is obtained after current annealing. Moreover, after subtracting best matching contact resistances at the two sides, $R_C=1.8$\,k$\Omega$ for the n-side and $R_C=2.1$\,k$\Omega$ for the p-side, the resistance peak is much more pronounced. This is seen in Fig.~S3(b). We point out here, that while the resistance now peaks quite strongly at the CNP, the pronounced photocurrent signal that appears at the CNP and is discussed in the main text appears sharper still. This points to the fact that the photocurrent is not simply proportional to the intrinsic resistance, but that relaxation processes will lead to a more subtle dependence on gate voltage.

After subtracting $R_C$ on both sides, the intrinsic graphene resistance end up in Fig.~S3(b) at $550$\,$\Omega$. However, the accuracy of this subtraction lies at around $100$\,$\Omega$, meaning that the intrinsic graphene resistance $R(0)$ at the CNP could range between $450$ and $550$\,$\Omega$. If we take $500$\,$\Omega$, we can get an estimate for the intrinsic conductivity $\sigma = R(0)^{-1}L/W$, where $L$ and $W$ are the width and length of the graphene sheet. We obtain $\sigma = 18$\,$e^2/h$. This is a large number, it is larger than the expected minimum conductivity of bilayer graphene. Taking two sheet, valley and spin degeneracy, $\sigma_{min} \leq 8$\, e$^2$/h. However, this theoretical minimum is the Sharvin resistance of the device, hence, a contact resistance which seemingly we have subtracted in part already. The fact that the experimentally deduced minimum $\sigma$ is so large supports the ballistic nature of our graphene device and a low puddle density in agreement with the measured low crossover density $n^*$. The later corresponds to an energy scale of $0.22$\,meV which in temperature corresponds to $2.5$\,K. Since the experiment has been conducted at a larger temperature of $8$\,K, it is not expected that the remaining puddles limit the conductance.